\documentclass[twocolumn,pre]{revtex4}
\usepackage{graphics}
\usepackage{graphicx}
\usepackage{amssymb}
\usepackage{amsmath}
\usepackage{hyperref}
\usepackage{float}
\usepackage{color}
\usepackage{color,soul}
\begin{document}

	\renewcommand{\thefootnote}{\fnsymbol{footnote}}
	
	\title{Size polydisperse model Ionic Liquid in bulk}

	\author{Somas Singh Urikhinbam}
	\author{Lenin S. Shagolsem}
	\email{slenin2001@gmail.com}
	\affiliation{Department of Physics, National Institute of Technology Manipur, Imphal, India} 
	
	\date{\today}

\begin{abstract}

\noindent The static and the dynamic properties of a size-polydisperse model ionic liquid is studied using molecular dynamics simulations. Here, size of the anions is derived from a Gaussian distribution while keeping cation size fixed, resulting in a system that closely corresponds to IL mixtures with a common cation. We systematically explore the behavior of thermodynamic transition temperatures, spatial ordering of ions and the resulting screening behavior as a function of polydispersity index, $\delta$. We observe a non-monotonic dependence of transition temperatures on $\delta$, and this non-monotonic behaviour is also reflected in other properties such as screening length. Furthermore, from the radial distribution function analysis it is found that, upon varying $\delta$, the spatial ordering of cations is affected, while no such changes is seen for anion. On the other hand, the analysis of ion motion through mean-square displacement show that for all $\delta$ values considered both inertial and diffusive regimes are observed (as expected in the liquid state). However, in contrast to neutral counterpart, the overall relaxation time of the polydisperse IL system increases (and hence decreasing diffusion coefficient) with increasing $\delta$.

\end{abstract}

\maketitle

\section{Introduction}
\label{sec: intro}

Ionic liquids (ILs) are salts consisting of molecular ions and they can exist in the liquid state at room temperature.\cite{freemantle} In contrast to simple molten inorganic salts such as NaCl and KCl, ILs exhibit interesting properties, e.g., low melting temperature, low vapour pressure, large electrochemical window, high thermal stability, high conductivity. Furthermore, the ability to modify properties through a proper choice of cation-anion pair make ILs a desirable solvent to work with, and therefore they are generally referred to as the `designer solvent'.\cite{freemantle1998} The possibility to tune chemical properties makes ILs exciting solvents. Another simple way to tune the properties of ILs is through mixtures, which is an extension from the `designer solvent' concept. Here, two or more pure ILs are mixed together to create a variety of novel solvents.\cite{niedermeyer2012, chatel2014, clough2015, di2020} Even with a small number of ILs one can modify the characteristics of the mixture while avoiding the complexities of synthesis, the purifying difficulties, and the expensive material costs.\cite{gouveia2019} A mixture, if close to ideal, has the potential to precisely fine-tune the properties within a range established by the pure components. Otherwise, the properties could be well outside the established range.\cite{mathew2015} The study of IL mixtures is vast and at the same time the ambiguous field of research due to the zoo of possible combinations of cation-anion pairs results in new ILs and hence significantly larger number of IL mixtures.

While there has been a significant amount of research related to pure ILs, studies addressing the behavior of IL mixtures are very limited. Mixtures of ILs have potential applications, e.g., as electrolytes in dye-sensitized solar cells and in liquid-liquid extraction, where the features of constituent ILs are combined to improve their performance.\cite{xi, garcia, garcia2} Other applications include gas solubility,\cite{finotello} solvent reaction media,\cite{kho} and as a gas chromatography stationary phase.\cite{baltazar} However, most of the research are primarily focused on binary mixtures, with only a few investigations on higher order mixtures.\cite{Siimenson2015} IL mixtures with organic solvents are also studied.\cite{thompson2019, watanabe2018} An observation which is consistent across a variety of binary mixtures is that random distribution of ions caused by Coulombic interactions dominate the structure of IL mixtures.\cite{payal2013, andanson2011, lui2011, abbott2011} A study on mixtures containing three components (1-ethyl-3-methylimidazolium tetrafluoroborate, 1-ethyl-3-methylimidazolium trifluoromethanesulfonate and 1-ethyl-3-methylimidazolium iodide) is reported in reference~\cite{Siimenson2015}. It is important to point out that in multi-component mixtures the constituent molecules differ in shape/size and energetic properties. It is the aspect of size disparity that we focus in this article (while keeping the remaining parameters fixed). Ideally, in the limit of large number of species, it is reasonable to assume a Gaussian distribution of size.  Polydispersity in size is common in glass former and supercooled liquids, colloidal fluids and dispersions, etc. The aspect of energy polydispersity is also studied in the context of model heteropolymers, biological fluids, high entropy alloys.\cite{lenin2015, lenin2016, rabin2019,vilip2022}

In this work, we consider a model IL with size polydisperse ions mimicking a multi-component mixture and investigate  its effect on thermodynamics, thermal-hysteresis behaviour, ordering, and dynamics as a function of polydispersity. Here, for simplicity, we assume the size distribution in anion only (and keep the cation size fixed). It is important to note that the model system considered in this study is an oversimplification of real ILs mixture, but we hope to gain insight on the phase and the order behavior, and the dynamics of the system in general. Since we consider a generic model, size distribution in cation (while keeping anion size fixed) would lead to the same effect except for the opposite charge polarity. Furthermore, to the best of our knowledge, no attempt has been made so far in addressing such extreme limit of multi-component mixture in the context of IL, and we hope that the results obtained here will serve as guide in making IL mixtures with large number of components. To this end, we employ a coarse-grained (CG) model of IL with charged spherical particles. Although simple CG model of ILs lacks detailed features of a real molecule/systems it can give important insights pertaining the static and the dynamic properties of the system in general. On the other hand, CG models have been extensively used in the study of ILs in fluid phase, where, for instance, it reproduces melting temperature less than 373 K (or 100$^{\circ}$C) corresponding to typical room temperature ILs by employing charged spherical model.\cite{gkagkas} Moreover, other researchers also employ similar CG model to investigate the effect of ion size asymmetry and short-range correlation on the electrical double layer of IL.\cite{fedorov} For example, see references \cite{capozza,fedorov2,fajardo} for the use of CG model in studying ILs.

The remainder of the paper is organized as follows. In section \ref{sec: model-description}, we describe the model and the simulation details, while in section \ref{sec: results} the results and the discussions are presented, and we conclude the paper in section \ref{sec: summary}. 



\section{Model and simulation details}
\label{sec: model-description}

Ionic liquid is modeled using generic coarse-grained spheres representing cation (C) and anion (A) molecules. Here, all the ions interact via Lennard-Jones (LJ) and Coulombic potential combined, i.e., 
\begin{equation}
	\label{eq: pair-potential}
	U(r_{ij}) = 4\epsilon_{ij}\left[\left(\frac{\sigma_{ij}}{r_{ij}}\right)^{12}-\left(\frac{\sigma_{ij}}{r_{ij}}\right)^{6}\right] + \frac{q_iq_j}{4\pi \epsilon_0\epsilon_rr_{ij}}~,
\end{equation}
with $\epsilon_{ij}$ as the LJ interaction strength, $r_{ij}$ as the separation between a pair of ions $i$-$j$, and $\sigma_{ij}=(\sigma_i+\sigma_j)/2$ is the effective particle separation between a pair of ions with $\sigma_{i/j}$ as the diameter of particle $i/j$. All the pair-wise interaction strength is set to $\epsilon_{ij}=0.03$ Kcal/mol. In this model, a single charge site is located at the ion center roughly corresponding to a monoatomic ion system. Note that in this relatively simple model interchanging the charge (i.e. making cation bigger and anion smaller) would leave the Hamiltonian invariant. 
On the other hand, studying molecular ion system would require an accurate description, for instance, charge distribution within a molecule has to be take into account which one model through charge arm description (i.e., different ion and charge centers).\cite{lindenberg2015,lu2016,lu2018}  

The charge on anion is fixed at $q_\text{\tiny A} = -e$, while that of cation is set to $q_\text{\tiny C}=+e$ where $e = 1.6 \times 10^{-19}$ C, and the dielectric constant of the medium $\epsilon_r=2$. The parameter values used here are the same as those reported in references.\cite{gkagkas1,gkagkas2,rupp,fajardo,capozza} Here, the mean size of anion ($\bar{\sigma}_\text{\tiny A} = 10$ \AA) is twice that of the cation (whose size is fixed at $\sigma_\text{\tiny C} = 5$ \AA). An example of system with anion larger than cation is $\left[ \text{EMIm} \right]^+\left[\text{TFSI}\right]^-$.\cite{aaron2018} 
In this study, we assume size distribution only in anion whose size is drawn randomly from Gaussian distribution, $P(\sigma_\text{\tiny A})= \frac{1}{\sqrt{2\pi(\text{SD})^2}} \exp \left[-\frac{1}{2 (\text{SD})^2} \left(\sigma_\text{\tiny A}-\bar{\sigma}_\text{\tiny A}\right)^2\right]$ with SD as the standard deviation, and $\bar{\sigma}_\text{\tiny A}$ as the mean. We consider different samples of varying polydispersity characterized by the polydispersity index, $\delta = \text{SD}/\bar{\sigma}_\text{\tiny A}$.

\begin{figure}[ht]
	\begin{center}
		\includegraphics[width=0.47\textwidth]{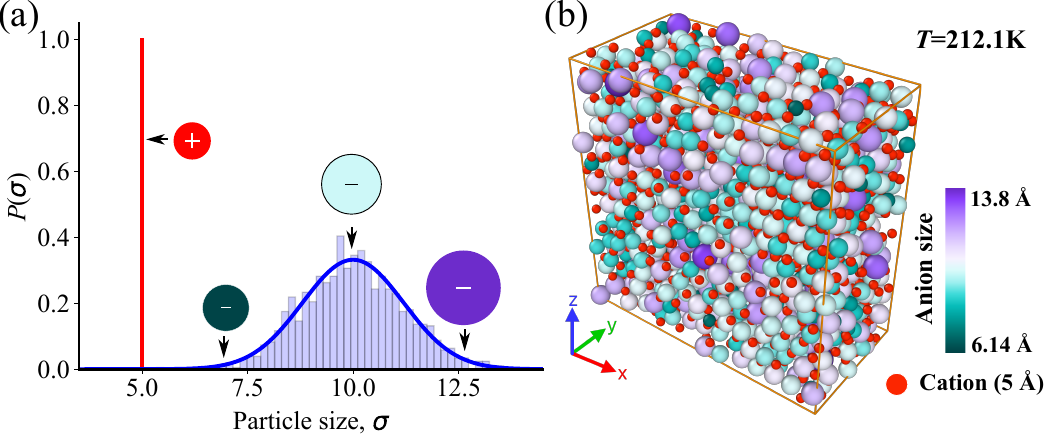} 
	\end{center}
	\caption{(a) Probability distribution with variance 0 (red) for cation size and 1.2 (blue) for anion size following Gaussian distribution, (b) Simulation snapshot of size polydisperse system with polydispersity index $\delta = 12\%$ in liquid phase. Color bar represents the size of the anions.}
	\label{fig: distribution}
\end{figure}

To investigate the size-polydisperse IL model system constant NPT molecular dynamics (MD) simulations \cite{allen} are carried out using open source simulation package LAMMPS\cite{lammps}. We use Nos\'{e}-Hoover thermostat and barostat to maintain constant temperature and pressure of the system, respectively. And we use the particle-particle particle-mesh (PPPM) solver for the slowly decaying Coulombic potential as it is shown to be a faster alternative to other methods.\cite{allen,frenkel,pollock} The equations of motion are integrated using velocity-Verlet algorithm with a time-step of 1 fs.

In this study, we consider 10 IL samples, viz., nine size-polydisperse systems with $0.3\% \le \delta \le 12\%$ and a sample with $\delta=0\%$ (i.e., system with no anion size variation and thus $\sigma_A=$10 \AA,~$\sigma_C=$5 \AA) as reference system. Each system consists of $N=2500$ ions in total with equal number of anion and cation and thus electro-neutrality is maintained in all the systems. 
Further, we also simulate the neutral counterpart (obtained by setting charge of each ion to zero) under pure athermal condition alongside for a qualitative comparison w.r.t. spatial distribution and dynamics. Since our interest in considering neutral system is to compare the trend under the variation of size polydispersity, we use LJ reduced units for neutral system, while atomic units are used for ionic systems.

The initial samples are prepared by placing ions in a simulation box (rectangular and periodic in all directions) which are then well equilibrated at high temperature under 1 atm pressure (where the systems are in disordered/liquid state). Typically we equilibrate the samples for 10 ns at a given temperature. Using the disordered liquid state, we perform production runs for various calculations.

\section{Results and discussion}
\label{sec: results}

Shown in figure \ref{fig: distribution}(b) is a simulation snapshot of a well equilibrated system at a relatively high temperature ($T=212.1$ K) for IL sample with $\delta=12\%$. As expected the spatial positions of ions are random, a typical feature of the liquid state. In the following, we discuss the effect of polydispersity on the melting/freezing of the IL systems through thermal cycles and locate the transition temperature which will be used as a reference point to compare among the different samples. 

\subsection{Thermodynamic melting temperature}
\label{subsec: thermalhysteresis}

\begin{figure}[h]
	\includegraphics[width=0.47\textwidth]{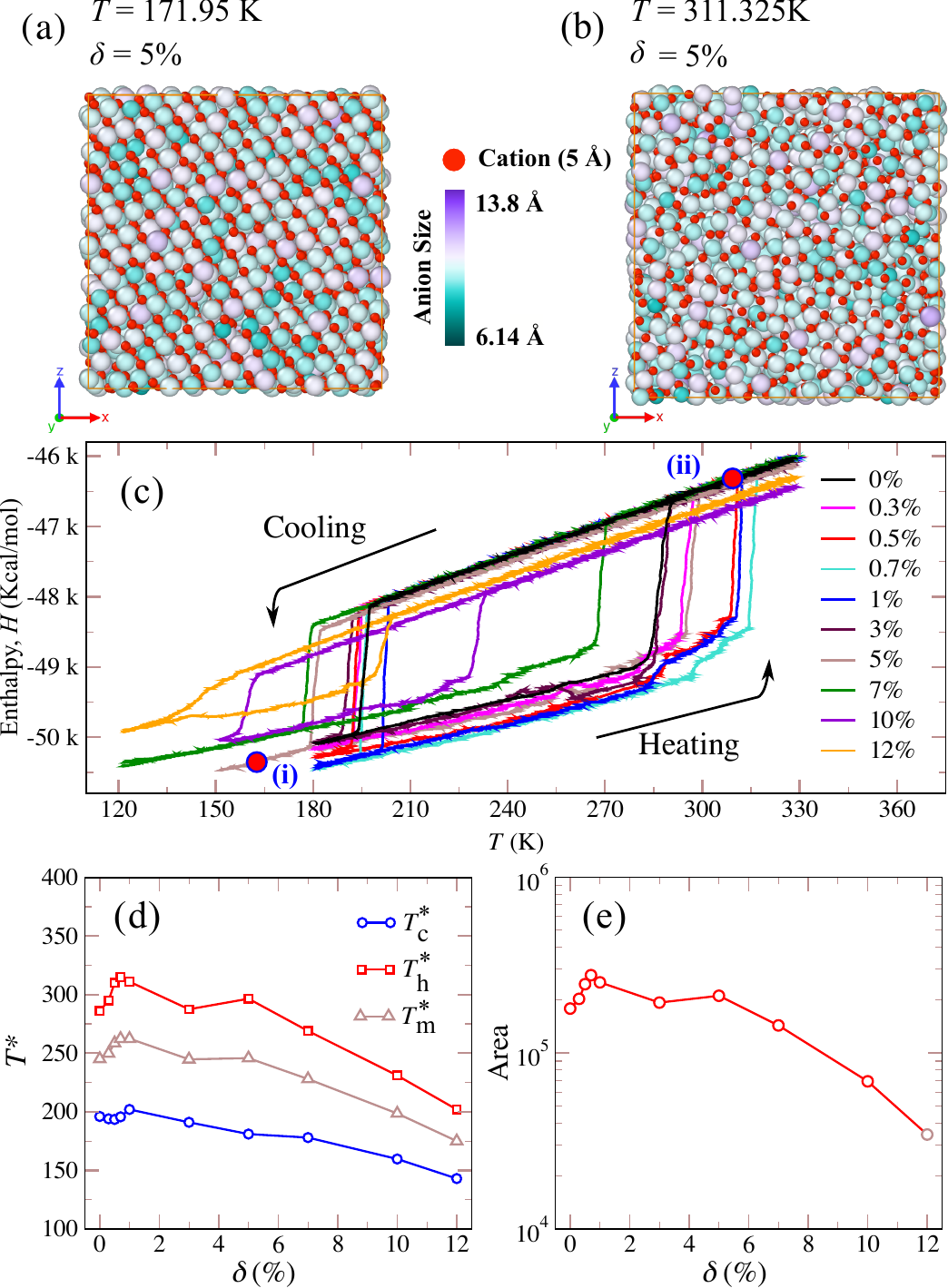}
	\caption{Simulation snapshot of system with $\delta=5\%$ in (a) the ordered phase (solid) at $T=171.95$ K, (b) the disordered phase (liquid) at $T = 311.325$ K, (c) Enthalpy, $H$ vs temperature, $T$ with different colours representing different polydispersity index indicated in legend, (d) Transition temperatures vs polydspersity index, $\delta$, showing an initial increase and decrease afterwards (e) Hysteresis loop area vs polydispersity index, $\delta$ showing the same trend as transition temperature.}
	\label{fig: hysteresis}
\end{figure}

\begin{figure*}[!ht]
	\includegraphics[width=0.9\textwidth]{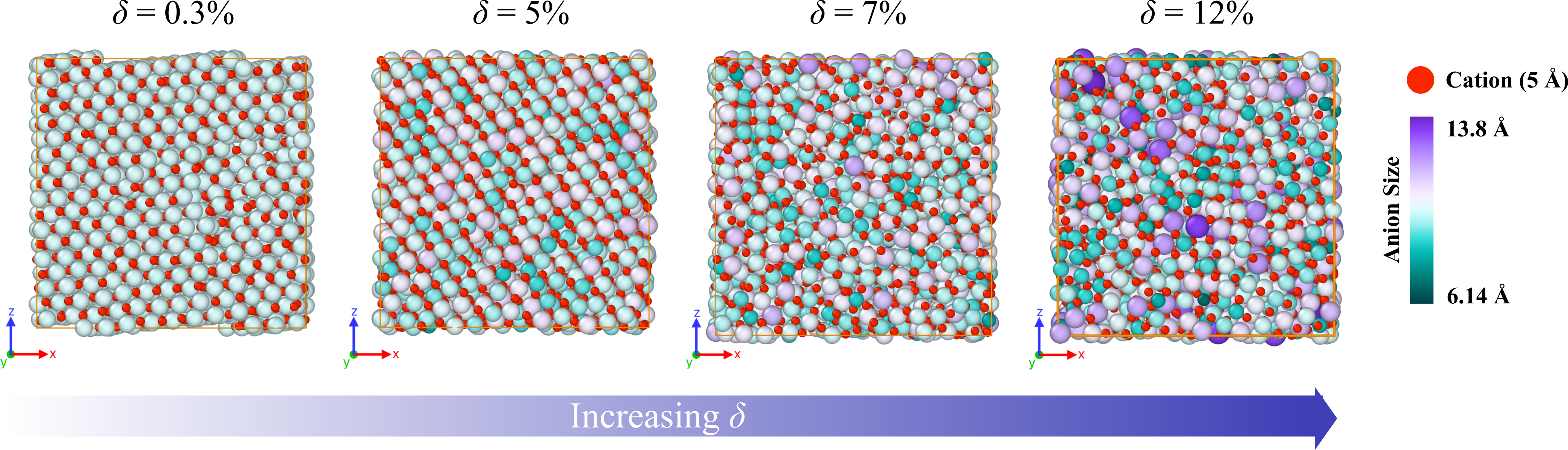}
	\caption{Equilibrium configuration of the charged system at different values of $\delta$ indicated in the figure (shown for temperature 5\% below the respective $T^*_c$, i.e. a frozen state).}
	\label{fig:solid}
\end{figure*}

From a disordered state at high temperature, the system is cooled down continuously to a sufficiently low temperature where it exhibits regular arrangement of particles followed by continuous heating up to the  same initial temperature, e.g. see figure~\ref{fig: hysteresis}(a)-(b). Thermal cycles are performed for all the systems at the temperature scan rate of $\lvert \text{d}T \rvert/\text{d}t = 1.5$ K/ns. Earlier studies have shown that the considered scan rate gives a reasonable thermodynamic freezing/melting temperature.\cite{gkagkas2,somas2022} 
The change of enthalpy, $H$, during the cooling-heating process is shown in figure \ref{fig: hysteresis}(c) where a hysteresis loop appears over one complete cycle. Here, $H=U+PV$ with $U$ as the internal energy, $P$ as the pressure, and $V$ as the volume. It is interesting to note that with increasing polydispersity the hysteresis loop area decreases and shifts to the lower temperature range. 
As we can see in the figure~\ref{fig: hysteresis}(c), during cooling the enthalpy changes abruptly at a particular value of temperature, $T^*_c$, indicating liquid to solid transition. Such abrupt change is also observed during heating (but at a different temperature $T^*_h$) indicating melting. The transition temperatures are reflected as sharp peaks in the plot of specific heat capacity, $c_P = \text{d}H/\text{d}T$ (not shown here). 
The thermodynamic melting temperature $T^*_m$ is then estimated using $T^*_m = T^*_c + T^*_h - \sqrt{T^*_c~T^*_h}$.\cite{luo2004} The obtained value of $T^*_m$ along with $T^*_c$ and $T^*_h$ at different values of $\delta$ is shown in figure~\ref{fig: hysteresis}(d), and the corresponding change of hysteresis loop area in figure~\ref{fig: hysteresis}(e). As we can see in the figure, $T^*_m$ (also loop area) increases with increasing $\delta$ and the maximum is observed at $\delta\approx 1\%$, beyond which the values decreases. Increasing $\delta$ increases the combinatorial entropy which in turn increases the entropy change associated with the change in state between the ordered and the disordered. This leads to the decrease in transition temperatures.\cite{warren1998} Furthermore, we can also see from figure \ref{fig:solid} that for $\delta = 7\%$ and $12\%$, the system is unable to form crystalline structure and instead forms glassy structure. This is consistent with previous studies on polydispersity, which found that if $\delta$ of the system exceeds the terminal polydispersity, it is unable to crystallize and instead forms glass.\cite{sarkar2015}

Since our aim is to investigate various properties of the ionic system in liquid phase, we relax the samples at temperature $5\%$ above their respective $T^*_h$ so that the systems are indeed in the liquid state and at same distance from the respective $T^*_h$. This choice of temperatures allows us a better comparison of the systems in their liquid state as fixing the same temperature for all is not ideal due the varying hysteresis loop area as well as thermodynamic melting temperature. In the following we discuss the effect of size polydispersity on the behavior of ion distribution and its effect on the screening length. 


\subsection{Spatial ordering and screening length} 
\label{sec: Spatial-ordering}

\begin{figure}[!ht]
	\begin{center}
		\includegraphics[width=0.47\textwidth]{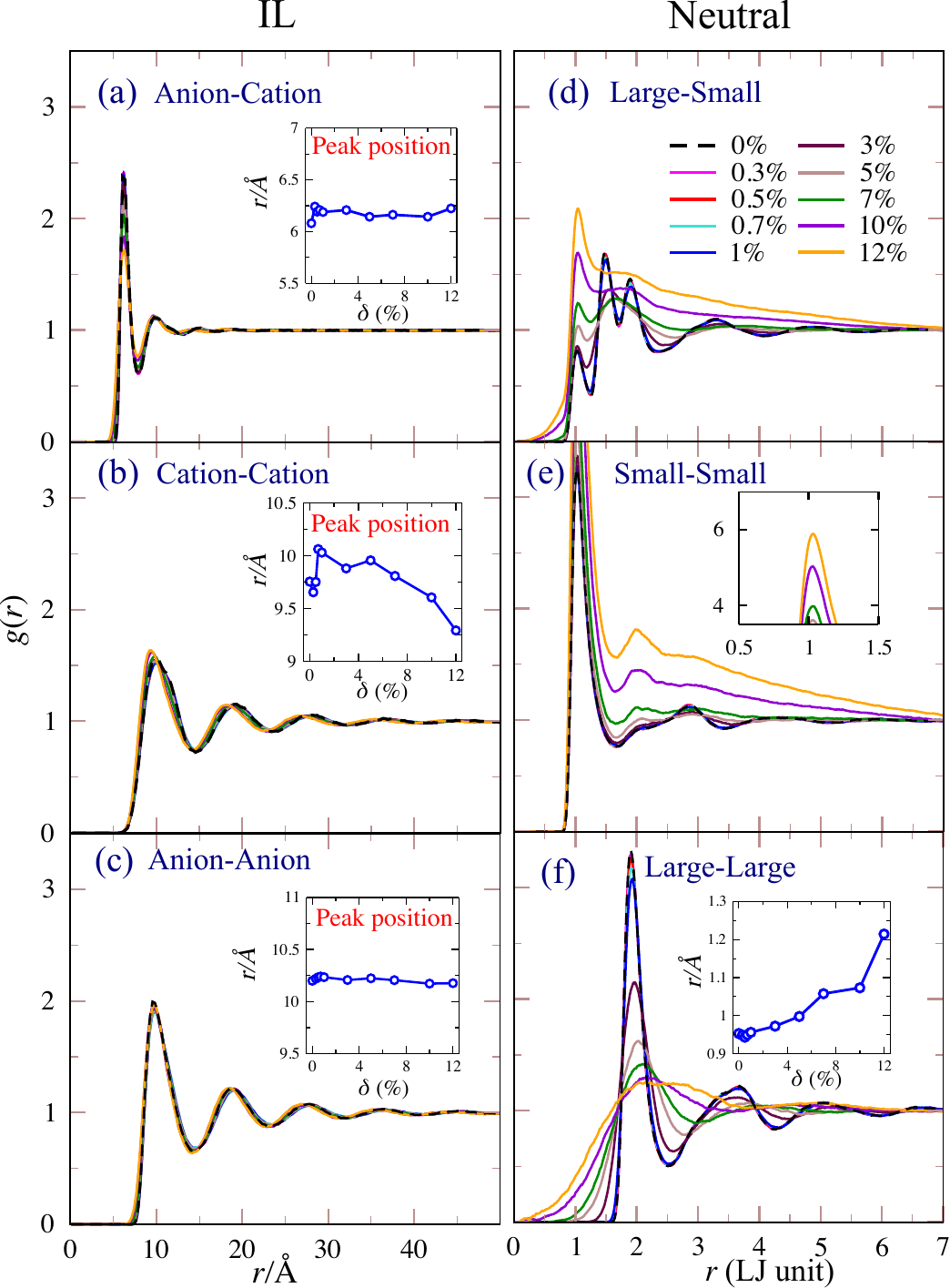}
		\caption{RDF as a function of normalized distance, $r/\sigma_{ij}$, for IL system (a-c), and its neutral counterpart (d-f) at different values of $\delta$ indicated in the figure. Inset of figures (a-c,f): First peak position of the respective $g(r)$ vs $\delta$, (e) cutoff part of the graph to make all the graph in the same y-range.} 
		\label{fig: rdf}
	\end{center}
\end{figure}

The spatial ordering of ion in the bulk is quantified through radial distribution function (RDF) defined as 
\begin{equation}
	\label{eq: rdf}
	g(r) = \frac{V}{N^2}\bigg \langle \sum_i \sum_{j \neq i} \delta \left(\textbf{r} - \textbf{r}_{ij} \right) \bigg \rangle
\end{equation}
with $\textbf{r}_{ij} = \textbf{r}_i - \textbf{r}_j$, $V$ and $N$ are volume and number of particles, respectively.\cite{allen} 
The RDF averaged over 500 independent configurations is shown in figure \ref{fig: rdf}(a)-(c) for IL systems, and in figure \ref{fig: rdf}(d)-(f) for the corresponding neutral systems. 

For the charged system, the overall $g(r)$ which represents an averaged over all the different species show no change in the peaks positions (except for decreasing amplitude) with increasing polydispersity, see figure \ref{fig: rdf}(a). It is clear from the figure that the mean closest approach, i.e., first peak position  $r_1 \approx 6.33$ \AA, and no significant changes observed upon increasing $\delta$. 
Because the interaction of like charged ions is repulsive in nature, they are not allowed to approach close to one other. As a result, the features of the first peak are dictated by the distance of closest approach between two oppositely charged ions. Increasing $\delta$ increases the variation in size of anion in the sample while maintaining the mean size constant. The same position of the first peak indicates that any two neighbouring ions in the system are separated by a fixed distance irrespective of their species. Combination of various sized ions results in shorter as well as longer pair separation distances, increasing peak width and in turn decreasing peak height. In the LJ systems, figure \ref{fig: rdf}(d), lack of Coulombic interaction resulted in the first three peaks at lower $\delta(<1\%)$ which correspond to the distance of closest approach between small-small, small-large and large-large particles consecutively. In large $\delta(>1\%)$, the second and third peak merged together due to the combinations of size dispersed particles.

In the Cation-Cation $g(r)$, figure \ref{fig: rdf}(b), first peak shifts to smaller $r$ as $\delta$ increases. The peaks are located at a distance roughly corresponding to twice the cation diameter ($\sigma_{\text{\tiny C}}$) which suggests the availability of space between cations that can be explored by anions.  As $\delta$ increases, smaller anions (ions with sizes at the lower end of the distribution) are introduced and they are able to readily reach the area between cations, and this lowers the effective repulsive force between cations, which is not achievable at smaller $\delta$. This leads to decrease in pair separation distance between cations, shifting the peak to smaller $r$ as $\delta$ increases. This shows the strong influence anion size has on the Cation-Cation ordering. For the case of neutral counterpart, the peak positions of different $\delta$ are at the same distance ($\sigma_{\text{\small s}}$), also the peak height increases and become narrower. This suggests that the size disparity of Large particles has no influence on the ordering of Small particles.
Comparing the RDFs of cation-cation and Small-Small, we notice that the first peaks for Cation-Cation is located at twice the cation diameter while that of Small-Small is located exactly at the diameter of small particle. This is expected because of the absence of Coulombic repulsion allowing the distance of closest approach to be $\sigma_{\text{\small s}}$.
In figure \ref{fig: rdf}(c), the Anion-Anion $g(r)$ shows no noticeable shift in the first peak. Despite having disparity in the size of anions, all the anions are separated by repulsive Coulombic force at a fixed distance, $r=\bar{\sigma}_{\text{\tiny A}}$. The amount of charge carried by the ion play a major role in maintaining the constant separation distance despite changing the ion size and cations has negligible influence on the ordering of anions. Where as in case of neutral system, the first peak of Large-Large $g(r)$ shift to the right which is due to the lack of Coulombic interaction and due to the variation in sizes of Large particles resulted to flattening and shifting of the peak to such degree, which are also observed in one component size-polydisperse systems.\cite{elena2018}

\begin{figure}[!h]
	\begin{center}
		\includegraphics[width=0.46\textwidth]{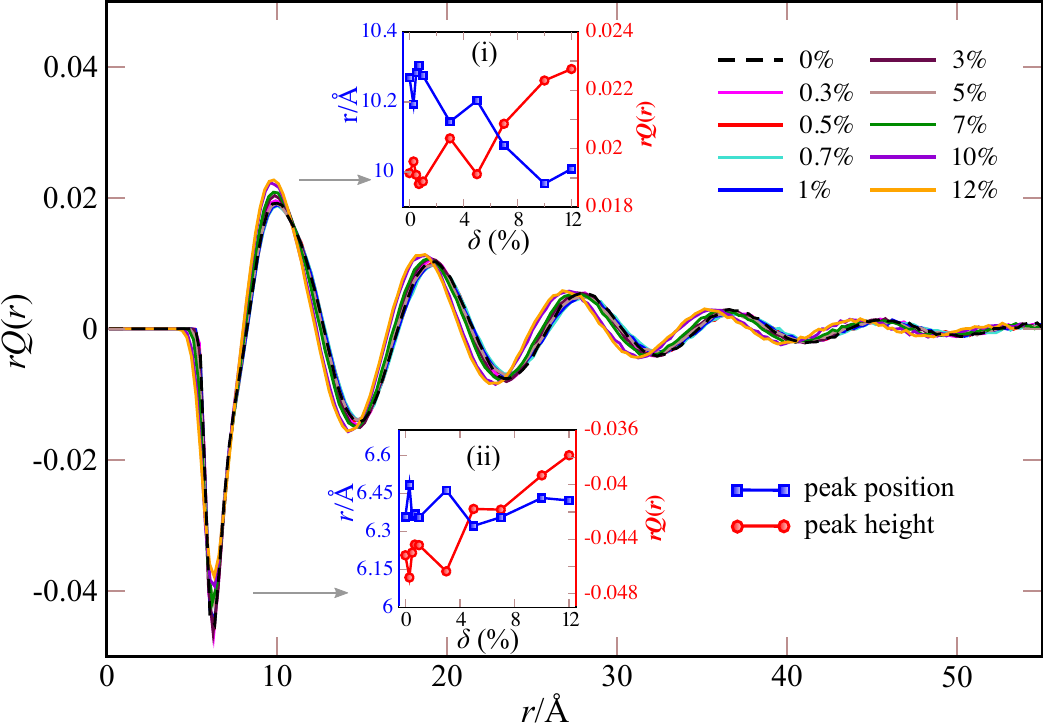}
	\end{center}
		\caption{Radial charge distribution plot for various $\delta$. Inset: variation of first (i) peak and (ii) anti-peak position (blue) and height (red) with $\delta$.}
		\label{fig: charge oscillation} 
\end{figure}

To have a better understanding of ion distribution in the bulk we also calculate the radial charge distribution function $Q(r)$ defined as
\begin{equation}
	\label{eq: rcd}
	Q(r) = \rho[g_{++}(r) + g_{--}(r) - 2g_{+-}(r)]~,
\end{equation}
with $\rho$ the average number density, $g_{++}(r)$ and $g_{--}(r)$ are cation-cation and anion-anion RDFs, respectively.\cite{keblinski} The charge distribution plotted as $rQ(r)$ vs $r$ for different polydisperse systems is show in figure \ref{fig: charge oscillation}. The observed exponentially screened oscillatory decay (persisting up to a distance $r \approx 10\sigma_C$) behaviour with alternating charge sign in each subsequent coordination shell indicates a strongly coupled ionic system. 
As shown in the insets of figure \ref{fig: charge oscillation}, with increasing $\delta$, the peak and anti-peak positions are shifted to lower $r$, also peak amplitude decreases (note that the shift is minor in the first anti-peak). 
\begin{figure}[!h]
	\begin{center}
		\includegraphics[width=0.47\textwidth]{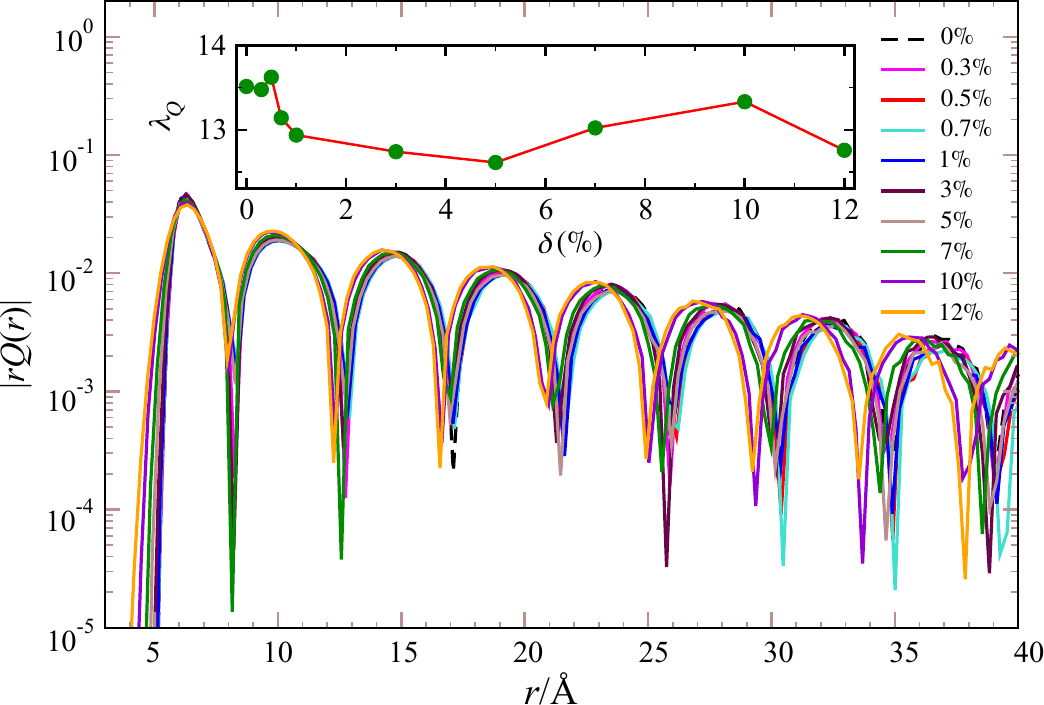}
		\caption{Plot of $\left| r Q(r) \right|$ for systems with different size polydispersity index $\delta$ indicated in the figure. Inset: The screening length $\lambda_Q$ obtained from the slope of the envelop of $\left| r Q(r) \right|-$plot shown as a function of $\delta$.}
		\label{fig: Screening Length} 
	\end{center}
\end{figure}

\begin{figure*}[!ht]
	\begin{center} 
		\includegraphics[width=0.9\textwidth]{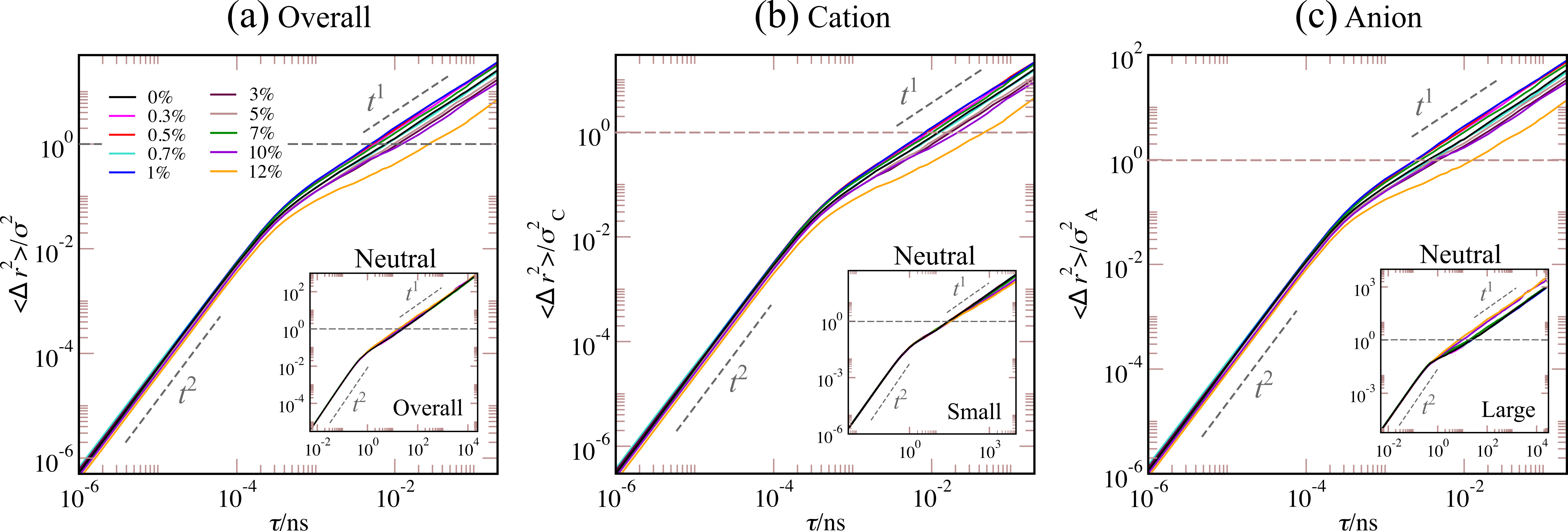}
		\caption{ MSD plot for the IL systems for (a) Overall (b) Cation (c) Anion showing both inertial and diffusive regimes. Inset: MSD for the corresponding neutral systems. The intersection of the horizontal dashed line at $\left< \Delta r^2 (t) \right>/\sigma_i^2 = 1$ with the MSD curves gives the mean relaxation time $\tau$.}
		\label{fig: msd} 
	\end{center}
\end{figure*}

It is well known that screening phenomena occurs in charges systems, e.g.~molten salts and ionic gases/liquids, which is the system's tendency to be locally charge neutral, and screening length $\lambda_Q$ is qualitatively understood as the characteristic length scale over which the system achieves local charge neutrality. Quantitatively, it is defined as the correlation length of radial charge distribution function. For weakly-coupled systems, well described by Debye-Huckel theory, $\lambda_Q$ is related to $Q(r)$ as $Q(r)\sim \frac{1}{r}\exp(-r/{\lambda_Q})$ which is a monotonically decaying function. While in the case of strongly-coupled systems $Q(r)\sim \frac{1}{r}\exp(-r/{\lambda_Q})\sin(2\pi r/d+\phi)$, which is oscillatory. Here, $d$ and $\phi$ are the period of oscillation and phase shift. The screening length can be extracted from the plot of $\left| r Q(r) \right|$ curve (plotted in semi-log scale) where the slope of the envelop of the curve gives $1/\lambda_Q$.\cite{keblinski2000}
The plot of $\left| r Q(r) \right|$ (along with $\lambda_Q$ as inset) is shown in figure~\ref{fig: Screening Length}, where we see that for the reference system (i.e.~$\delta=0\%$) $\lambda_Q\approx13.5$ \AA~(which is roughly $1.35~\bar{\sigma}_{\text{\tiny A}}$). With increasing $\delta$ the value of $\lambda_Q$ increases slightly (maximum at $\delta\approx0.5\%$) and then decreases with minimum at $\delta\approx5\%$ (where $\lambda_Q\approx12.6$ \AA, a decrease of roughly 6.7\% w.r.t.~reference system) which is again followed by an increase and decrease. 
From the cation-cation RDF, see figure~\ref{fig: rdf}(b), it is clear that the observed changes in $\lambda_Q$ is a consequence of the rearrangement of cations in the presence of size disperse anions. And with increasing $\delta$ the local Coulombic field is screened more effectively (due to the presence of smaller anions) and thus highlights the possibility of tuning screening length only using size as a parameter. 
In the following section we discuss the effect that size polydispersity produces on the dynamics of ions.


\subsection{Dynamics} \label{sec: dynamics}

The dynamics of ions are quantified through mean square displacement (MSD) defined as
\begin{equation}
	\label{eqn:msd}
	\left< \Delta r^2 (t) \right> = \frac{1}{N} \sum^N_{i=1} \left[r_i(t) - r_i(0)\right]^2~,
\end{equation}
where $r_i(t)$ is the position of $i^{\rm th}$ particle at time $t$, and $N$ the number of particles. From which the diffusion constant $D_0$ is obtained as 
\begin{eqnarray}
D_0 = \lim_{t\rightarrow \infty} \frac{\langle \Delta r^2(t) \rangle}{6t}.
\label{eqn: diffusion-coeff} 
\end{eqnarray}
In general, $\left< \Delta r^2 (t) \right> \sim t^\alpha$, where the value of exponent $\alpha$ determines the qualitative nature of the particle motion, i.e. when $\alpha = 1$ the motion is diffusive, and for $\alpha \ne 1$ it shows anomalous diffusion.\cite{burov2011} 
In figure~\ref{fig: msd}, we display MSD curves for both charged and corresponding neutral systems at different values of $\delta$. In all the systems, inertial regime (i.e.~$\Delta r^2 \sim t^2$ at short time scales) as well as diffusive regime (i.e.~$\Delta r^2 \sim t^1$ at longer time scales) are observed and thus the evolution of particle dynamics is same across the systems. However, the consequences of varying particle size distribution through $\delta$ is reflected in the relaxation time and hence the diffusion constant. Here, the relaxation time, $\tau$, is defined as the average time for a particle to cover its own size $\sigma_i$ (which in the case of particle type with size polydispersity we consider $\sigma_i=\overline{\sigma}$ mean size) and thus $\tau$ represents the longest relaxation time of the system. 

\begin{figure}[!ht]
	\begin{center} 
		\includegraphics[width=0.47\textwidth]{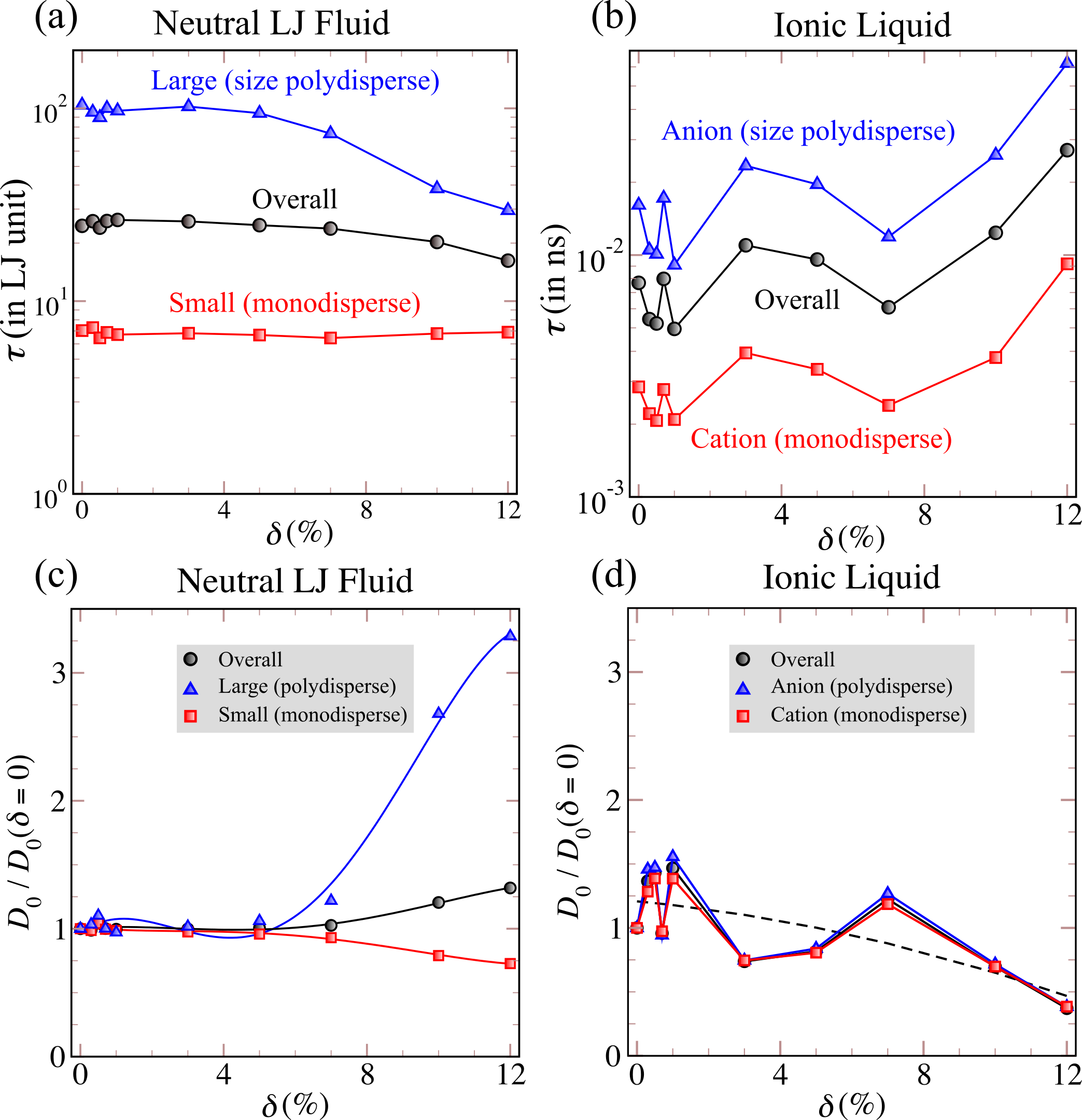}
		\caption{Relaxation time, $\tau$ vs $\delta$ for (a) neutral system, and (b) IL, and the corresponding diffusion constant are shown in (c) and (d), respectively. Solid lines in figure (c) are fit using polynomial equation of order 4 (corresponding to equation~\ref{eqn: diffusion const}), while the trend lines in figure (d) are fit using quadratic equation.}
		\label{fig: relaxation time} 
	\end{center}
\end{figure}

In figure~\ref{fig: relaxation time}(a) and (b), the relaxation time as function of polydispersity index is shown for the reference neutral system and IL, respectively. For neutral system, it is observed that, as $\delta$ increases, the overall value of $\tau$ decreases and hence $D_0$ increases, see figure~\ref{fig: relaxation time}(d). 

Comparison between the two types of neutral particles (i.e.~large size-polydisperse and small monodisperse) reveal that $\tau$ decreases for large particles, while for small type it remains constant and overall the trend is dictated by that of the large size-polydisperse particle. 

The decrease of $\tau$ can be understood from the fact that with increasing $\delta$ we introduce particles whose size are smaller and larger than the mean size and it is the smaller ones whose relaxation time is small and effectively it dominates. Consequently, with increasing $\delta$, the diffusion constant $D_0$ increases rapidly for large, while a slight decreases is observed for small particle, an opposite trend, and overall $D_0$ increases. The observed increase of $D_0$ with $\delta$ for large size-polydisperse particles is consistent with other analytical and simulations works,\cite{prem2022,malvern2019,xu1988} where the normalized diffusion constant is shown to vary as
\begin{equation} \label{eqn: diffusion const}
\frac{D}{D_{\rm ref}} \approx 1-a\frac{\delta^2}{4}+b\frac{\delta^4}{32} + O(\delta^6)~,
\end{equation}
where $a$ and $b$ are constants, and $D_{\rm ref}$ the diffusion constant for reference monodisperse system. 
It is interesting to note within the system there is competing dynamics and the behavior of small monodisperse particles dictates the overall dynamics. This observation further suggest the possibility to tune $D_0$ and hence the viscosity of the sample with varying particle fraction in addition to that of polydispersity index as parameter.

On the other hand, for IL systems, the overall value of $\tau$ increases (non-monotonically) as $\delta$ increases, where it is larger for the anion (i.e.~larger ion with size distribution) compared to that of the cation and hence $D_0$ decreases, see figure~\ref{fig: relaxation time}(c). Of the smaller and larger particles introduced with the increase in $\delta$, smaller particles exhibit stronger Coulombic interaction due to their small distance of closest approach. Since Coulombic interaction follows inverse square law, the interaction between two oppositely charged ions becomes stronger as the distance between them is reduced. As $\delta$ increases, this is the case for the interaction between cation and the anions whose size are smaller than the mean size. This strong Coulombic interaction inhibits mobility of both the ion species which in turn increases their relaxation times, resulting in the decrease in diffusion coefficient. The overall $\tau$ is dominated by the size-polydisperse anions as seem in figure \ref{fig: relaxation time}(b).

The average diffusion coefficient of the system as well as that of the respective constituent ions shows the same behaviour on increasing $\delta$. A quadratic polynomial fit shows the trend line, see dashed lines in figure \ref{fig: relaxation time}(d). Thus, introduction of smaller particles reduces the diffusion coefficients of all the particles in the Coulombic system showing that the dynamic behaviour of the system can be tuned using the size of one species as a parameter.




\section{Conclusion}
\label{sec: summary}

In this work, we have studied the influence of ion size distribution on the static and dynamics properties of a model ionic liquid system. We assume the size distribution only in anions while the size of cations remains constant. This corresponds to an IL mixture with large components which shares the same cation.

Studying the transition temperature through thermal cycles we found that the hysteresis loop area as well as thermodynamic melting temperature have a non-monotonic dependence on polydispersity index $\delta$. An initial increase is observed with increasing $\delta$ upto $1\%$ and then decreases afterwards. This non-monotonic trend is consistent across other properties such as spatial ordering, screening length and relaxation time.
The spatial ordering analysis further reveal that the size disparity of anion dictates the ordering of cations. Smaller sized anions, that are introduced on increasing $\delta$, can easily access the regions between cations and it decreases the net repulsive interaction thereby decreasing the separation distance between cations. On the other hand, this disparity in size of anions did not affect its spatial arrangement. This rearrangement of cations in the presence of small anions affects the screening length $\lambda_Q$. The value of $\lambda_Q$ increases slightly with maximum at $\delta=5\%$ and then decreases with minimum at $\delta=5\%$ which is followed by an increase and then decrease.

Furthermore, the average ion relaxation time is found to increases with increasing $\delta$. Coulombic interaction hindering the ion mobility causes the increase in relaxation time which results in the decrease of diffusion constant. This is a contrasting feature as compared to that of the neutral counterpart whose relaxation time decreases (in turn increases diffusion constant) as $\delta$ increases.
In conclusion, we have studied a size-polydisperse model IL in bulk and highlight the possibility to tune the static and dynamic properties of IL using size as a controlled parameter. Another interesting aspect, from the application point of view, is the behavior near an electrified interface which will be discussed in an upcoming work.

\begin{acknowledgements}
S.S.U. acknowledges the financial support from National Institute of Technology Manipur. Fruitful discussions with M.~Birla is gratefully acknowledged. 
\end{acknowledgements}

\end{document}